\documentclass[pdftex,twocolumn,epjc3]{svjour3}     
\pdfoutput=1

\usepackage{verbatim}
\usepackage{listings}
\RequirePackage[T1]{fontenc}
\usepackage{shadow}
\usepackage{varioref}
\usepackage{array}
\usepackage{rotating}
\usepackage{multirow}
\usepackage{hhline}

\smartqed  

\usepackage{floatrow}
\usepackage[strict]{changepage}

\usepackage{myaasmacros}

\RequirePackage{graphicx}
\RequirePackage{mathptmx}      
\RequirePackage{flushend}
\RequirePackage[numbers,sort&compress]{natbib}
\RequirePackage[colorlinks,citecolor=blue,urlcolor=blue,linkcolor=blue]{hyperref}

\usepackage{amsmath}  
\usepackage{amssymb}  
\usepackage{subfig}  
\usepackage{graphicx}  
\usepackage{paralist}  
\usepackage{multirow}  
\usepackage{amsbsy}  
\usepackage{amsfonts}  
\usepackage{cuted}  
\usepackage{xspace}  
\PassOptionsToPackage{numbers,sort&compress}{natbib}  
\usepackage{lmodern}  

\journalname{Eur. Phys. J. C}

\begin{document}

\title{Possibility of hypothetical stable micro black hole production at future 100 TeV collider}

\author{A.V. Sokolov\thanksref{a,b,c,e1}
        \and
        M.S. Pshirkov\thanksref{d,a,e} 
}

\thankstext{e1}{e-mail: anton.sokolov@physics.msu.ru}

\institute{Institute for Nuclear Research of the Russian Academy of Sciences, 117312, Moscow, Russia\label{a}
\and
Physics Department, Lomonosov Moscow State University, Moscow 119991, Russia\label{b}
\and
Institute of Theoretical and Experimental Physics, B.Cheremushkinskaya 25, Moscow 117218, Russia\label{c}
\and
Sternberg Astronomical Institute, Lomonosov Moscow State University, Universitetsky prospekt 13, 119234, Moscow, Russia\label{d}
\and
P.N. Lebedev Physical Institute, Pushchino Radio Astronomy Observatory, 142290 Pushchino, Russia\label{e}}


\date{Received: date / Accepted: date}

\maketitle

\begin{abstract}
We study the phenomenology of TeV-scale black holes predicted in theories with large extra dimensions, under the further assumption that they are absolutely stable. Our goal is to present an exhaustive analysis of safety of the proposed 100 TeV collider, as it was done in the case of the LHC. We consider the theories with different number of extra dimensions and identify those for which a possible accretion to macroscopic size would have timescales shorter than the lifetime of the Solar system. We calculate the cross sections of the black hole production at the proposed 100 TeV collider, the fraction of the black holes trapped inside the Earth and the resulting rate of capture inside the Earth via an improved method. We study the astrophysical consequences of stable micro black holes existence, in particular its influence on the stability of white dwarfs and neutron stars. We obtain constraints for the previously unexplored range of higher-dimensional Planck mass values. Several astrophysical scenarios of the micro black hole production, which were not considered before, are taken into account. Finally, using the astrophysical constraints we consider the implications for future 100 TeV terrestrial experiments. We exclude the possibility of the charged stable micro black holes production.
\end{abstract}





\section{Introduction}

Unsolved puzzles of fundamental physics encourage scientists to probe interactions at progressively  higher  energies. The Large Hadron Collider has not so far found any hints on  'new physics', so there are plans to construct even more  energetic and luminous experiment. In particular, there is the High-Luminosity LHC project \cite{1} that aims to increase the LHC luminosity by a factor of seven and the Future Circular Collider project \cite{2} that can ultimately reach the hadron collision energy of 100~TeV.


Before the launch of the LHC the question of its safety was examined in detail by the LHC Safety Assessment Group (e.g. see review \cite{Ellis:2008hg}). It was shown that the hypothetical exotic kinds of matter, such as strangelets, magnetic monopoles, true vacuum bubbles and stable micro black holes, cannot be produced at the LHC, for their existence at the energies below 14~TeV strongly contradicts astrophysical observations. In this work we study the case of the future 100~TeV collider and limit our research to the hypothetical stable micro black holes. The case of 100~TeV collider differs significantly from the case of the LHC, because relevant astrophysical scenarios are considerably altered at progressively higher energies.

Microscopic black holes with the masses under 100~TeV can naturally appear in the extra-dimensional theories \cite{ArkaniHamed:1998rs,Antoniadis:1998ig,Randall:1999ee}. In these theories the value of the higher-dimensional Panck mass can be as low as several TeV. Existing constraints on the parameters of the models with extra dimensions come from the direct measurements of Newton's law of gravitation at small distances \cite{Adelberger:2009zz,Murata:2014nra} and from the LHC searches \cite{Khachatryan:2014rra,Aad:2015zva} for the missing transverse energy of jets due to graviton-emission processes: the radii of the extra dimensions $R_D < 37~\mu$m and the Planck mass $M_D > 3.5~$TeV. In this work we will conservatively assume $M_D > 3~$TeV. The minimum black hole mass corresponding to the certain $M_D$ can be found from the condition that the entropy of the black hole be large (see \cite{Giddings:2001bu}). In the case of six dimensions the black hole mass $M=5M_D$ corresponds to the entropy $S_{BH} \simeq 24$ and we will take it as the lower threshold of mass.

In this work we assume that there is no Hawking radiation \cite{Hawking:1974sw}. This scenario should be taken into account due to the lack of the experimental data concerning the Hawking radiation as well as due to the theoretical uncertainties in the field of quantum gravity. However, one should keep in mind that the black hole evaporation is an inevitable consequence of quantum theory. Even though there are theoretical suggestions \cite{Unruh:2004zk,Vilkovisky:2005db} that the process of Hawking radiation could depend on the details of the Planck-scale degrees of freedom, once the black hole acquires enough mass to enter a semiclassical regime, the universal Hawking radiation starts. The timescale of evaporation is faster than that of mass acquisition, so the black hole cannot grow macroscopically according to any theoretical considerations.
Based on the data from astrophysical observations, our work conducts a test of safety, independent of the reliance on theoretical results.

The produced micro black holes, while in general having non-zero charge, can either lose their charge immediately via the Schwinger mechanism \cite{Schwinger:1951nm} or remain charged. On the one hand, there is a similarity between the Hawking radiation and the Schwinger mechanism, studied in many works, e.g. \cite{Srinivasan:1998ty, Kim:2007qu, Ellis:2008hg}. This suggests that in the hypothetical case of the absence of the Hawking radiation the Schwinger discharge can be absent as well and the micro black holes would remain charged. On the other hand, there is also a difference between these effects: the Hawking radiation, unlike the Schwinger mechanism, is a trans-horizon effect and the conversion of vacuum fluctuations into particles in this case does not occur over a well-defined space-time domain. If we assume that it is the horizon physics that, despite all the theoretical evidence, forbids the Hawking evaporation, then the Schwinger mechanism can still operate, hence the neutrality of the black holes. For the sake of robustness we consider the both cases: the neutral stable micro black holes as well as the charged ones.


In our work we refer often to the methods proposed in the study \cite{Giddings:2008gr} of safety of the LHC in the context of the stable micro black holes production. However, we propose an improved method of the calculation of the number of the black holes trapped inside the Earth during the work of high energy collider. Besides, we examine some astrophysical mechanisms of micro black hole production, that were not considered in \cite{Giddings:2008gr} and which provide conservative model-independent constraints.

\section{Accretion times}

This section mainly reviews and structurizes the results of the article \cite{Giddings:2008gr} about the higher-dimensional black hole accretion inside the Earth in order to justify the choice of the gravitational theories we consider in the following sections. 

Let us consider the D-dimensional gravitational action with a general compact metric $g_{mn}(y)$ and a warp factor $A$:
\begin{equation}
S=\frac{1}{8\pi G_D} \int{d^Dx\sqrt{-g}\cdot \frac{1}{2}R},
\end{equation}
\begin{equation}
ds^2 = e^{2A(y)}dx^{\mu}dx_{\mu} + g_{mn}(y)dy^mdy^n,
\end{equation} 
where $x^\mu $ are the usual non-compact Minkowski coordinates (here and below we use natural units $\hbar = c = k_B = 1$). The characteristic radius $R_D$ of the extra dimensions is connected with the Planck mass $M_D$ as follows:
\begin{equation}
\frac{M_4^2}{M_D^2} \simeq (R_D M_D)^{D-4} \cdot e^{2\Delta A},
\end{equation}
where $M_4 = 2.4\cdot 10^{15}~$ TeV, $~\Delta A$ is a difference in warping between the region with the maximum warp factor and the standard model region. In the case of zero warping and  $D=5$ the Planck mass $M_D \sim 10~$TeV gives the value $R_5 \sim 10^7 \thinspace$km, that is obviously excluded. In the case of the theories with the larger number of dimensions $R_D$ decreases and the value of $R_6 \sim 5 \thinspace  \mu m$ is already smaller than the existing constraint on $R_D$ ($R_D < 37~\mu$m).

The micro black hole accretion generally goes through three different phases: subnuclear, subatomic and macroscopic, that follow each other while the black hole is growing. In the cases of 1 or 2 extra dimensions the black hole initial capture radius in matter is larger than the nuclear size. For instance, in the case $D=6$ the capture radius is $R_{EM} \sim 10^{-12}~\text{cm} $. As a consequence, the accretion goes through the subatomic phase from the very beginning and there is no need to consider the subnuclear phase. In the case of the bigger number of extra dimensions we conservatively neglect the subnuclear phase, for it is sufficient to find the lower constraint on the black hole accretion time. 

The micro black hole accretion is the fastest in the theories without warping ($\Delta A = 0$), for the characteristic radius of extra dimensions in this case is maximal and the transition to the slow four-dimensional accretion regime occurs later. Thus, in order to find the lower constraint on the black hole accretion time, we consider the case $\Delta A = 0$, apart from the theory with $D=5$, where the warping is needed to meet the experimental constraints. In the latter case we set the warp factor value so that $R_5$ be maximal and equal to $37 \thinspace \mu$m.

\begin{figure*}[h!]
\caption{Lower bound on the Earth accretion time in the theories with $D$=6 dimensions and values of the Planck mass from 3 to 20~TeV}\label{1f}
\centering \includegraphics[height=6cm]{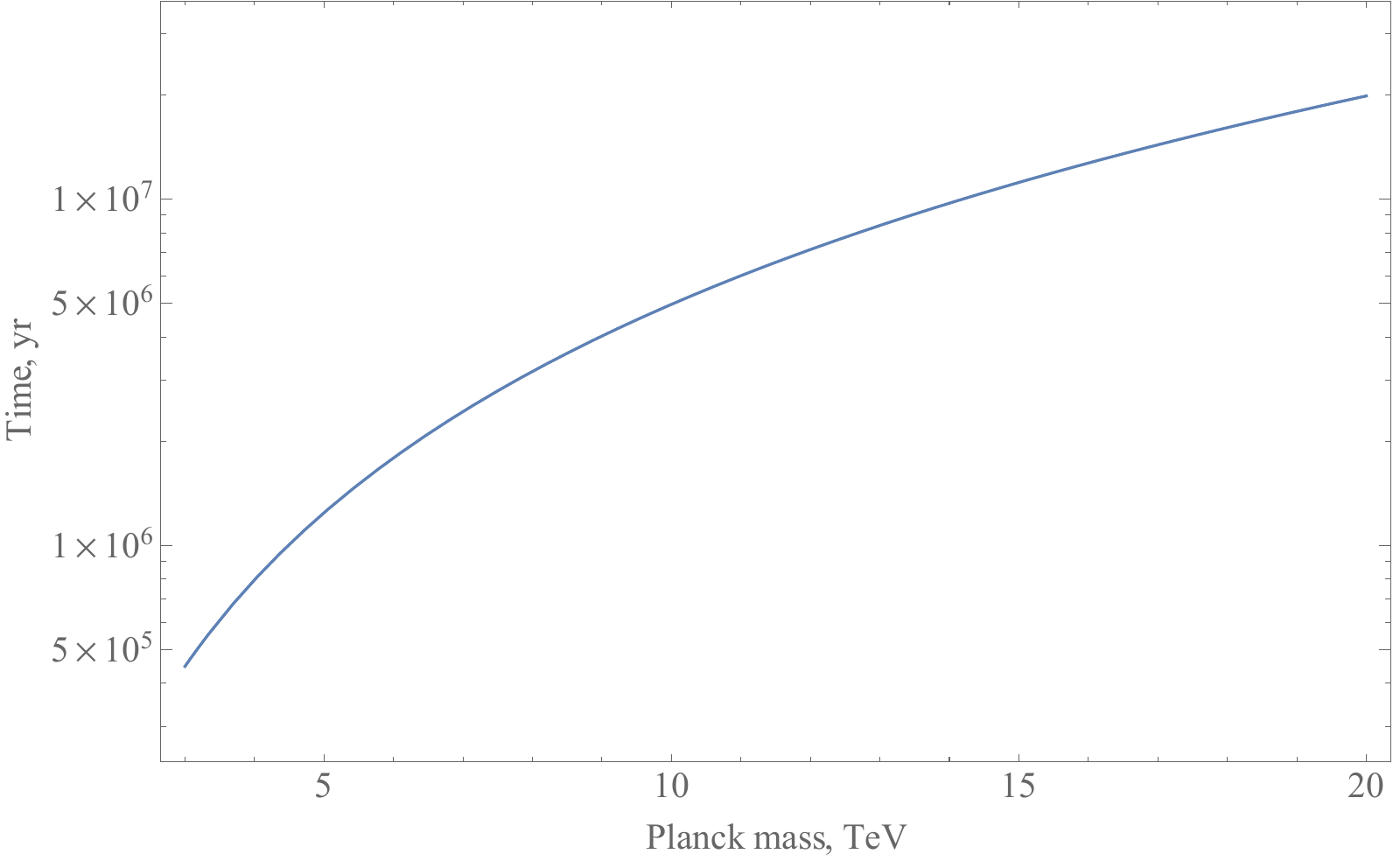}
\end{figure*}

The constraints on the times of the micro black hole accretion inside the Earth in the theories with $D=5-11$ are presented in Table \ref{1t}. The capture radius during the phase of macroscopic (Bondi) accretion is called Bondi radius $R_B$, while the radius $R_C$ denotes the distance of the crossover, where the higher-dimensional gravity force equals the four-dimensional one. The accretion times depend on the properties of the matter inside the Earth. One can parametrize this dependence by the Debye temperature $T_{Deb}$ (that is about 400\thinspace K for the materials typical for Earth's composition) and numerical $\mathcal{O}(1)$ constants $\chi $ and $\lambda_D$ ($4 \leqslant \lambda_4 < 18,\quad 3< \lambda_D <6.6$ in the case $D>4$), which slightly depend on the material. We see that in the theories with more than six dimensions the accretion times are larger than the lifetime of the Solar system. In the cases of 5 and 6 dimensions we get accretion times $t \gtrsim 10^5-10^6~$yr, which are not exceedingly large from the geological point of view, though the case $D=5$ requires a theory with the special choice of warping:
\begin{equation}\label{6}
19-1.5 \ln \frac{M_5}{M_0} < \Delta A < 24-1.5 \ln \frac{M_5}{M_0},
\end{equation}
where $M_0=1~$TeV and the limits are given by the experimental constraint on the value of $R_5$ and  the condition that the accretion time is shorter than the Solar system lifetime. The results for both 5 and 6 dimensions are similar, and, bearing in mind that the theory with $D=5$ requires an extreme fine-tuning, we limit our research to the case $D=6$. The full time of the black hole growth inside the Earth in this case is shown in Fig. \ref{1f}.
 
\begin{table*}[h!]
\caption{Accretion times of the stable micro black holes inside the Earth divided into subatomic (capture radius smaller than $a=$1~\AA) and macroscopic phases. The macroscopic phase is divided into three, division governed by the radius of extra dimensions $R_D$ and the crossover radius $R_C$. The last phase corresponds to the growth from $R_C$ to the size of the black hole with the mass comparable to the mass of the Earth.}\label{1t}
\resizebox{16.5cm}{4.2cm}{

\begin{tabular}{|c|c|c|c|c|} \hline
\multirow{2}{*}{\small{Phases}} & \multirow{2}{*}{\small{Subatomic}} & \multicolumn{3}{|c|}{\small{Macroscopic}} \\ \hhline{~~---}
&& $a \lesssim R_B \lesssim R_D$ & $R_D \lesssim R_B \lesssim R_C$ & $R_B > R_C $ \\ \hline
$D=5,$ & \multirow{2}{*}{$t \gtrsim 2.3\cdot 10^{-5} \cdot \frac{T_4^2}{\chi} m_5^3 $~s} & \multirow{2}{*}{$t \lesssim 3.6 \cdot 10^{-2} \frac{1}{\lambda_5} m_5^3 $~s} & \multirow{2}{*}{$t > 3\cdot 10^5$~yr} & \multirow{2}{*}{$t \thicksim 3\cdot 10^5$~yr} \\ 
 $\Delta A \thicksim f(M_5)$ &&&&\\ \hline
$D=6$ & $t \gtrsim 4.5\cdot 10^{3} \cdot \frac{T_4^2}{\chi} m_6^4 $ s & $t = 5.5 \cdot 10^{4} \frac{1}{\lambda_6} m_6^2 $ yr & $t = 2.4 \cdot 10^{5} \frac{1}{\lambda_6} m_6^2 $ yr & $t = 9.7 \cdot 10^{4} \frac{1}{\lambda_4} m_6^2$ yr \\ \hline
$D=7$ & $t \gtrsim 3.0\cdot 10^{11} \cdot \frac{T_4^2}{\chi}  m_7^5 $ s & $t = 8.6 \cdot 10^{8} \frac{1}{\lambda_7} m_7^{5/3} $ yr & $t = 9.1 \cdot 10^{9} \frac{1}{\lambda_7} m_7^{5/3} $ yr & $t = 1.3 \cdot 10^{10} \frac{1}{\lambda_4} m_7^{5/3}$ yr \\ \hline
$D=8$ & $t \gtrsim 5.4\cdot 10^{6} \cdot \frac{T_4^2}{\chi} m_8^{-3/2} $ yr $+ t' $ & \multicolumn{3}{|c|}{\multirow{4}{*}{$t = 1.2 \cdot 10^{12} \cdot \frac{1}{\lambda_4} $ yr \quad $(a < R_B < 2a)$}} \\ \hhline{--~~~}
$D=9$ & $t \gtrsim 2.0\cdot 10^{4} \cdot \frac{T_4^2}{\chi} m_9^{-7/5} $ yr $+ t' $ & \multicolumn{3}{|c|}{} \\ \hhline{--~~~}
$D=10$ & $t \gtrsim 2.2\cdot 10^{2} \cdot \frac{T_4^2}{\chi} m_{10}^{-4/3} $ yr $+ t' $ & \multicolumn{3}{|c|}{} \\ \hhline{--~~~}
$D=11$ & $t \gtrsim 4.8 \cdot \frac{T_4^2}{\chi} m_{11}^{-9/7} $ yr $+ t' $ & \multicolumn{3}{|c|}{} \\ \hline
\multicolumn{5}{l}{$f(M_5) = 19-1.5 \ln m_5$,~ $t' = 3.1 \cdot 10^{11} \cdot \frac{T_4^2}{\chi} $ yr,~ $m_D = M_D/M_0$,~ $M_0 = 1$~TeV,~ $T_4 = T_{Deb}/400$~K,~ $a = 1~$\AA,}\\
\multicolumn{5}{l}{$\chi \simeq 1$,~ $4 \leqslant \lambda_4 < 18,\quad 3< \lambda_D <6.6$,~ $\Delta A $ -- warp factor} \\ 
\end{tabular}
}
\end{table*}

\section{Production of gravitationally bound black holes}

\begin{figure*}[h!]
\caption{The number\protect\footnotemark of the black holes produced during the lifetime of the future 100~TeV collider with the integrated luminosity $L=10^4$~fb$^{-1}$~ in the theories with $D$=6 dimensions, Planck masses $3-15$~TeV and for two values of the inelasticity parameter $y=0.5$ and $y=0.7$}\label{2f}
\centering \includegraphics[height=6cm]{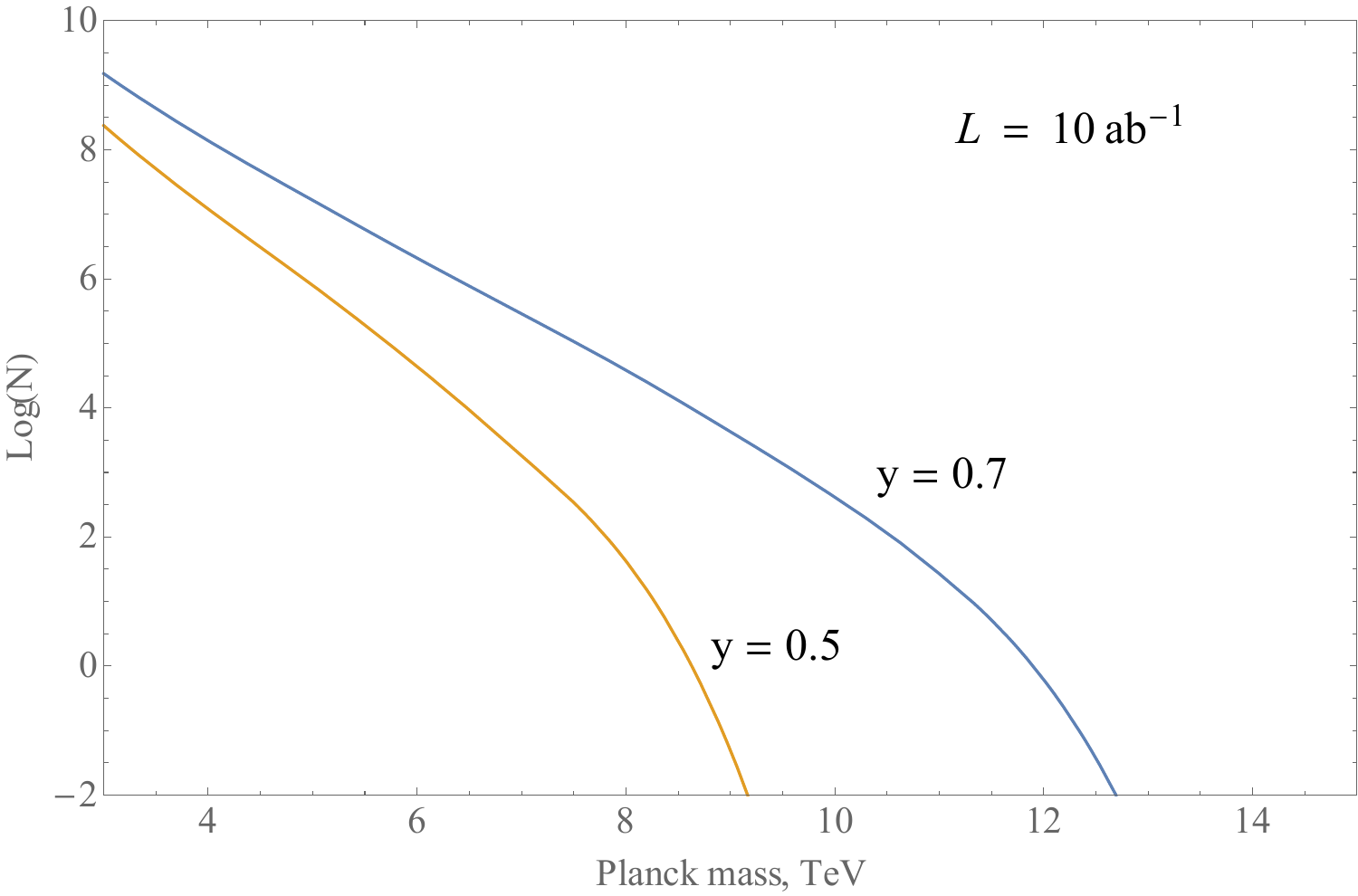}
\end{figure*}

The cross section of the black hole production in pp collisions at the energy of 100\thinspace TeV according to the factorization theorem \cite{Collins:1989gx} is:
\begin{equation}\label{2}
\sigma_{BH} (M>M_{min}) = \sum\limits_{ij} \int\limits_{\tau_{min}}^1 d\tau \int\limits_{\tau}^1 \frac{dx}{x} f_i(x)f_j(\tau /x)  \sigma '  (\sqrt{s'}),
\end{equation}
where $f_i(x)$ are parton distribution functions (we use the set CT14qed \cite{Schmidt:2015zda}),~ $\tau_{min} = \frac{M_{min}^2}{y^2s}$,~ $M_{min}=5M_6$,~ $\tau = x_1 x_2$,~ $s' = s \cdot \tau$, ~$y \simeq 0.5-0.7$ is the inelasticity factor (see \cite{Eardley:2002re}). The double sum implies the summation over the all pairs of partons. The cross section for the collision of two partons is:
\begin{equation}
\sigma '  (\sqrt{s'}) = \pi R^2(\sqrt{\tau s})/4, \quad R(\sqrt{\tau s}) = \frac{1}{M_6}\cdot \left( \frac{3\sqrt{\tau s}}{4M_6} \right) ^{1/3},
\end{equation}
$R$ is the Schwarzshild radius. We allow black hole production only for the partonic collisions with the impact parameter $b < 0.5 R$, following the work \cite{Giddings:2007nr}. The parton distribution functions should be taken at the scale $Q \sim 1/R$, as discussed in \cite{Giddings:2001bu}. The number of the black holes produced at the future collider (integrated luminosity $L \sim 10^4~$fb$^{-1}$, center of mass energy $\sqrt{s} = 100~$TeV,~ PDF scale $Q \sim 10\thinspace$TeV) is plotted as a function of $M_6$ in Fig. \ref{2f}.

\begin{figure*}[h!]
\begin{floatrow}
\ffigbox{\caption{The fraction of the neutral black holes produced at the future 100~TeV collider which are trapped inside the Earth (theories with 6 dimensions)}\label{3f}}
{\includegraphics[height=4cm, width=6cm]{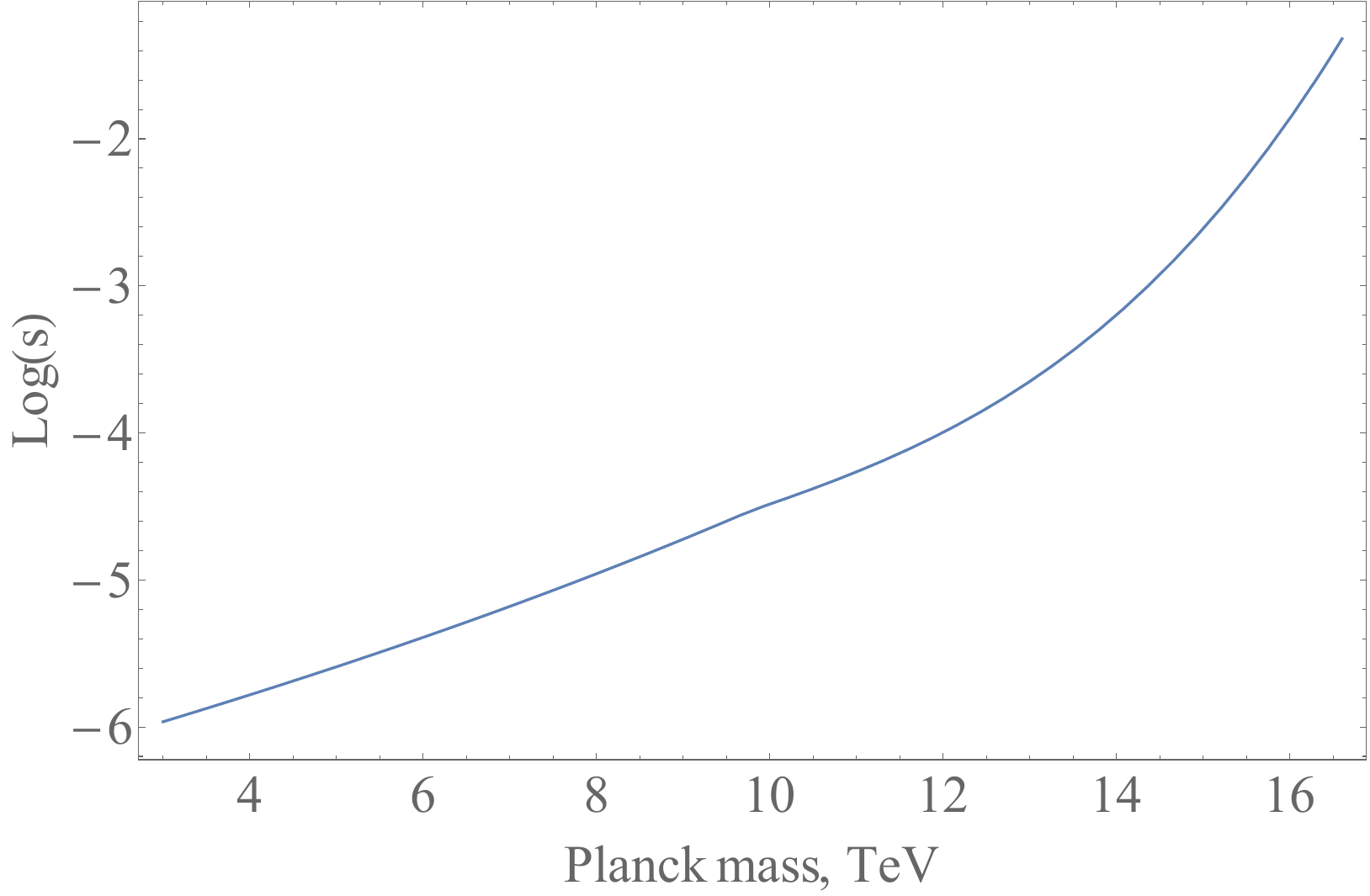}}
\ffigbox{\caption{The amount of the neutral black holes trapped inside the Earth during the lifetime of the future 100~TeV collider (integrated luminosity $L=10^4~$fb$^{-1}$) with the suppression factor Eq.~(\ref{eq:suppression-factor}) taken into account and for two values of the inelasticity $y=0.5$ and $y=0.7$}\label{4f}}
{\includegraphics[height=4cm, width=6cm]{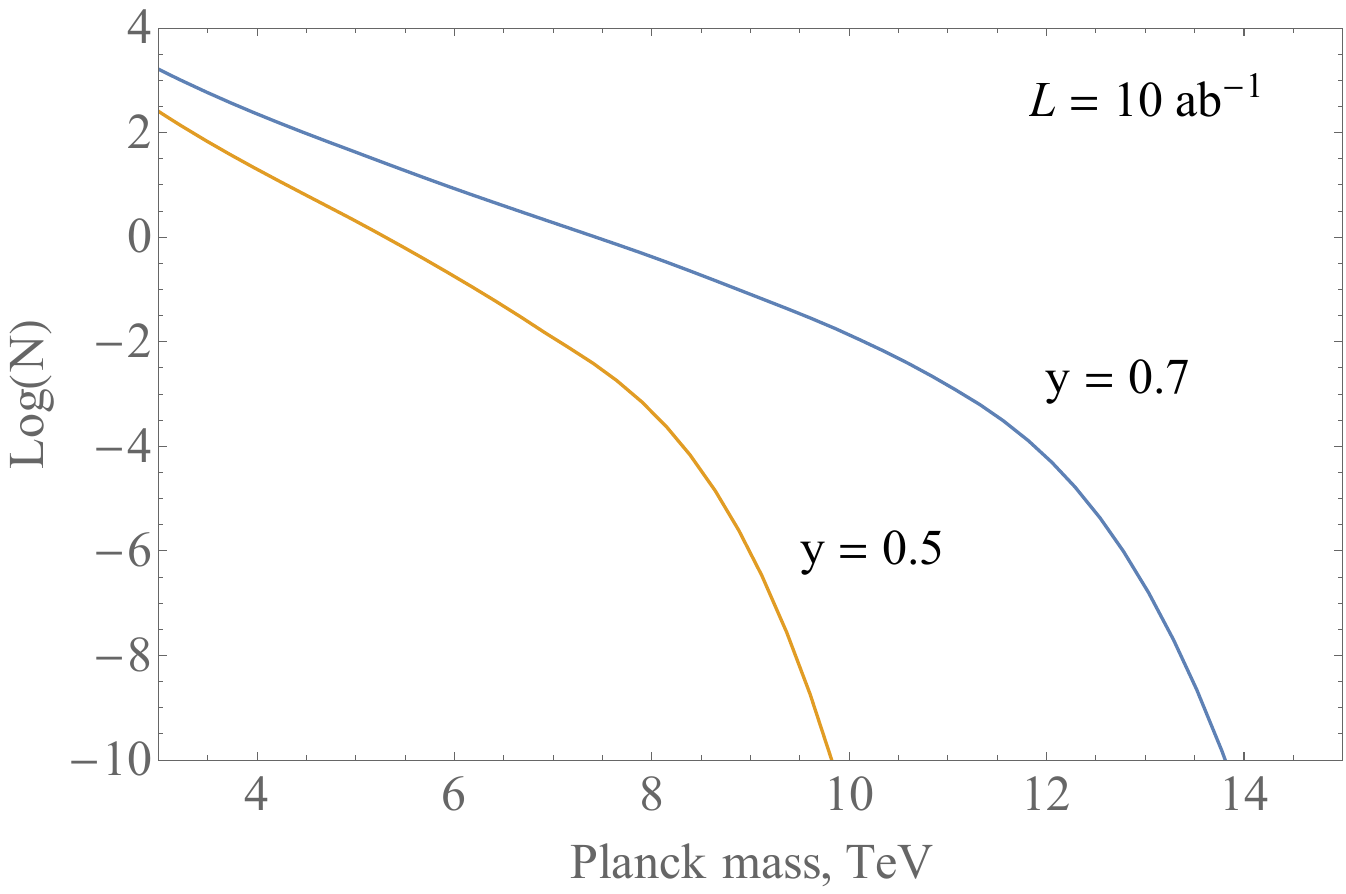}}
\end{floatrow}
\end{figure*}

Black holes on average will be produced with velocities much larger than the escape velocity, thus  only a tiny fraction of them will be trapped inside the Earth. In order to calculate this fraction we need the distribution of not only the longitudinal momenta of partons, given by the standard parton distribution functions, but also of the transverse momenta as well, that is why we use the transverse momentum dependent parton distribution functions (TMDPDF) $g_i(x,k)$ from the library tmdlib-1.0.7 \cite{Hautmann:2014kza}. According to the TMD factorization \cite{Rogers:2015sqa}, the cross section of the black hole production (its mass being larger than $M$, longitudinal momentum less than $p$ and transverse momentum less than $k$) is given by:
\begin{align}
\begin{split}
\sigma_{BH} (M,p,k) &= \frac{1}{2\pi}\sum\limits_{ij} \int\limits_0^{2\pi} d\alpha \times \\
 \int\limits_{R(p,k,\alpha)} & dk_1 dk_2 dx_1 dx_2 \cdot g_i(x_1,k_1)g_j(x_2,k_2)  \sigma (x_1,x_2),
\end{split}\\
\sigma (x_1,x_2) &= \frac{\pi}{4 M_6^2}\cdot \left( \frac{3\sqrt{x_1 x_2 s}}{4M_6} \right) ^{2/3},
\end{align}
\begin{equation}
\begin{split}
R(p,k,\alpha) = \left\lbrace \frac{M}{\sqrt{s}}\leq \right. & \left. x_1, x_2 \leq 1,~ \frac{\sqrt{s}}{2} |x_1-x_2| \leq p, \right. \\
& \left. \sqrt{k_1^2+k_2^2+2k_1 k_2 \cos\alpha} \leq k  \right\rbrace .
\end{split}
\end{equation}
At the present moment we will assume that the black holes are neutral. Neutral black holes will slow down inside the Earth due to the accretion and gravitational scattering. This process was in detail studied in \cite{Giddings:2008gr}, where the maximum speed  for a black hole to be still trapped inside the Earth was calculated. In our case ($D=6$):
\begin{equation}
v_{max} = 11.2\cdot 10^{-3} \cdot \frac{l}{d} \left( \frac{M_0}{M_6} \right) ^{8/3} \left( \frac{M_0}{M} \right) ^{1/3} ,
\end{equation}
where $l$ is the path length of the black hole inside the Earth, $d$ is the diameter of the Earth, $M_0=1~$TeV. In case this velocity is smaller than the escape velocity $v_E$ one should take the latter as a trapping threshold. Note that $l/d$ equals $v_r/v$, the ratio of the radial (directed to the center of the Earth) speed to the total speed of the black hole; $v_r = (k/M)\cos\phi$, where $\phi$ is the angle between the direction to the center of the Earth and the transverse momentum, $-\pi/2<\phi<\pi/2$, the velocities are non-relativistic. Then the condition for the black hole to be trapped reads as:
\begin{equation}
\label{1}
\begin{split}
v&< \max \left[ v_{max},v_{E} \right] \Rightarrow p^2+k^2 < \\
& \max \left[ 11.2\cdot 10^{-3} \left( \frac{M_0}{M_6} \right) ^{8/3} \left( \frac{M_0}{M} \right) ^{1/3} k M \cos\phi,\>  M^2 v_{E}^2 \right].
\end{split}
\end{equation}
Thus, the fraction of the black holes that will be trapped (suppression factor) is given by:
\begin{equation}
\label{eq:suppression-factor}
\begin{split}
s(M_6) = &\frac{1}{2\pi \sigma_{tot}(M_6)} \int\limits_{-\pi/2}^{\pi/2}d\phi \times \\
& \int\limits_{D(M_6,\phi)} \left| \frac{d^3 \sigma_{BH}(M_6)}{dpdkdM} \right| dpdkdM,
\end{split}
\end{equation} 
where the region $D(M_6,\phi)$ is given by the inequality (\ref{1}) and the total cross section $\sigma_{tot}(M_6)$ -- by the Eq. (\ref{2}). The values of the suppression factor are presented in Fig. \ref{3f} and the number of the black holes trapped inside the Earth for the integrated luminosity $L=10^4$~fb$^{-1}$ is plotted in Fig. \ref{4f}.

It is worth noting that the calculation of the suppression factor in Ref. \cite{Giddings:2008gr} was considerably simplified: the authors assumed that all the black holes which were produced had the same lowest possible mass $M_{min}$ and that their transverse momenta are identical. These assumptions lead to the understated value of the suppression factor, because the maximal initial momentum for a black hole to be still trapped inside the Earth is $p_{max} = \max \left[ Mv_{max}, Mv_E \right] \propto M^{2/3}$ or $M$ and, setting $M=M_{min}$, one gets the underestimated value of this momentum for the black holes with $M>M_{min}$. This means that not all the trapped black holes were taken into account. For example, in our case assumption that $M = M_{min}$ for all the black holes produced leads to the decrease in the number of the trapped black holes by the factor $0.5 - 0.7$ depending on the Planck mass.

\footnotetext{Here and below the logarithmic scale in plots is based on decimal logarithms}

\section{Astrophysical constraints}
\subsection{General considerations}
If the stable micro black holes could be produced at the colliders, they would be as well naturally produced in the Universe in the interactions of high energy cosmic rays. This can contradict the observed long lifetimes of the dense astrophysical
objects, in which these black holes can get stuck and accrete. First of all, we would like to consider the stopping power of  different astrophysical objects. The theory of the deceleration of the microscopic black holes inside celestial bodies was developed in \cite{Giddings:2008gr}. It was calculated that the Earth has not sufficient power to stop neither neutral nor charged (with masses more than 7~TeV) relativistic micro black holes (we consider the black holes produced in the collisions of the high energy cosmic rays with some slow moving particles), while the Sun can stop the charged relativistic black holes with the masses well in excess of 100 TeV and cannot stop the neutral relativistic black holes. Finally, the general expression for the minimum column density required to stop the neutral micro black hole with the mass $M$ was derived. In our case ($D=6$) this required column density is:
\begin{equation}
\delta_{min} = 0.27 \cdot \left( \frac{M_6}{M_0} \right) ^3 \left( \frac{\gamma_i M}{M_6} \right) ^{1/3} M_0^3,
\end{equation}
where $M_0^3 = 4.6\cdot 10^{12}~$g/cm$^2$,~ $\gamma_i$ -- the initial Lorentz factor of the black hole. Consider a cosmic ray nucleus with atomic number $A$ and high energy $E$ hitting a target nucleon. The initial energy of the produced black hole $\gamma_i M =y x E/A $ ($y$ is the inelasticity parameter, $x$ is  the fraction of the nucleon centre of mass momentum carried by the incident parton). Then the condition for the column density $\delta$ required to stop the black hole is:
\begin{equation}\label{3}
\delta > \delta_{min} \Rightarrow xE < \left( \frac{\delta}{0.27M_0^3} \right) ^3 \left( \frac{M_0}{M_6} \right) ^9 \cdot M_6 \cdot \frac{A}{y}.
\end{equation}
The column density along the diameter of a white dwarf with the mass $M_{WD}=1.2M_{\odot}$ is $\delta_{WD} = 3.8\cdot 10^{16}~$g/cm$^2$, the same value for a neutron star is $\delta_{NS} \sim 10^{20}~$g/cm$^2$. Due to the inequality (\ref{3}), a neutron star can stop practically all the black holes going through it: $E$ is limited from above by more than $10^8~$TeV. For such energies the cosmic ray flux is negligible. That is not the case for the white dwarfs, so we have to account for the condition (\ref{3}) in the further calculations of the white dwarf constraints.

There are several mechanisms that could provide significant fluxes of the micro black holes. The most efficient mechanism is the collision of the high energy cosmic rays with the surfaces of  dense stars. However, considering this mechanism, one has to account for the large magnetic fields of these stars, which presence can lead to the considerable synchrotron energy losses of the cosmic rays. This question was considered by \cite{Giddings:2008gr} (see Appendix G there). It was shown that the magnetic screening prevents the cosmic rays with the energies more than $E_{max}$ from reaching the surface,
\begin{equation}\label{4}
E_{max} \simeq 1.8\cdot 10^{17} \text{eV}\cdot \frac{A^4}{Z^4} \frac{10 \text{km}}{R} \left( \frac{10^8 \text{G}}{B\sin\theta} \right) ^2,
\end{equation}
where $Z$ is the charge of the cosmic ray nucleus, $R$ is the star's radius, and  $\theta$ is the angle between the momentum of the cosmic ray and the magnetic axis of the star. The synchrotron losses are negligible in the case of the ordinary white dwarfs with the polar magnetic field $B \sim 10^5~$G ($E_{max} \gtrsim 10^{20}~$eV), but they play an important role in the case of neutron stars, for which the smallest detected magnetic field is $B \simeq 7\cdot 10^7~$G \cite{Manchester:2004bp}.

\subsection{Constraints from white dwarfs}

The flux of the micro black holes produced by the cosmic rays hitting the white dwarf surface is:
\begin{equation}\label{5}
\begin{split}
\phi_{BH} = b & \int\limits_{E_{min}}^{E_{max}} A(E) J(E) dE \times \\
& \sum\limits_{ij} \int\limits_{\tau_{min}}^1 d\tau \int\limits_{\tau}^1 \frac{dx}{x} f_i(x)f_j(\tau /x)  \sigma '  (\sqrt{s'}),
\end{split}
\end{equation}
where $b=1/\sigma_{NN}$,~ $\sigma_{NN}=100~$mb -- total nucleon-nucleon inelastic cross section,~ $E_{min} = \min[M_{min}^2A /$ $(2m_p y^2)] = 7\cdot 10^5~$TeV,~ $E_{max} = 2\cdot 10^8~ \text{TeV}$,~ $y = \max[0.5,$ $~ M_{min}/$ $100~\text{TeV}]$ (most conservative case, corresponding to the minimum production at the 100~TeV collider),~ $\tau_{min} = M_{min}^2A /$ $(2m_p y^2E)$,~ $J(E)$ -- the combined energy spectrum of cosmic rays as measured by the Auger Observatory, fitted with a flux model (see \cite{Aab:2015bza}). The dependence of the mean atomic number $A$ on the energy $E$ is taken from \cite{Aab:2015bza} as well: we interpolated Auger data and averaged over the two hadronic interaction models EPOS-LHC and QGS-JetII-04. However in the case of the white dwarf the resulting bounds weakly depend on the cosmic ray composition: the black hole production at the given energy decreases with increasing $A$, but also the condition (\ref{3}) becomes less constraining and allows higher energies to be involved.

\begin{figure*}
\begin{floatrow}
\ffigbox{\caption{The amount of the neutral black holes, stopped by a white dwarf with a radius 5600~km during $10^9$ yr for the different cosmic ray composition: $100\%$ p, $50\%$ p, $10\%$ p fractions and corresponding to Auger data on the mean atomic number}\label{5f}}
{\includegraphics[height=4cm, width=6cm]{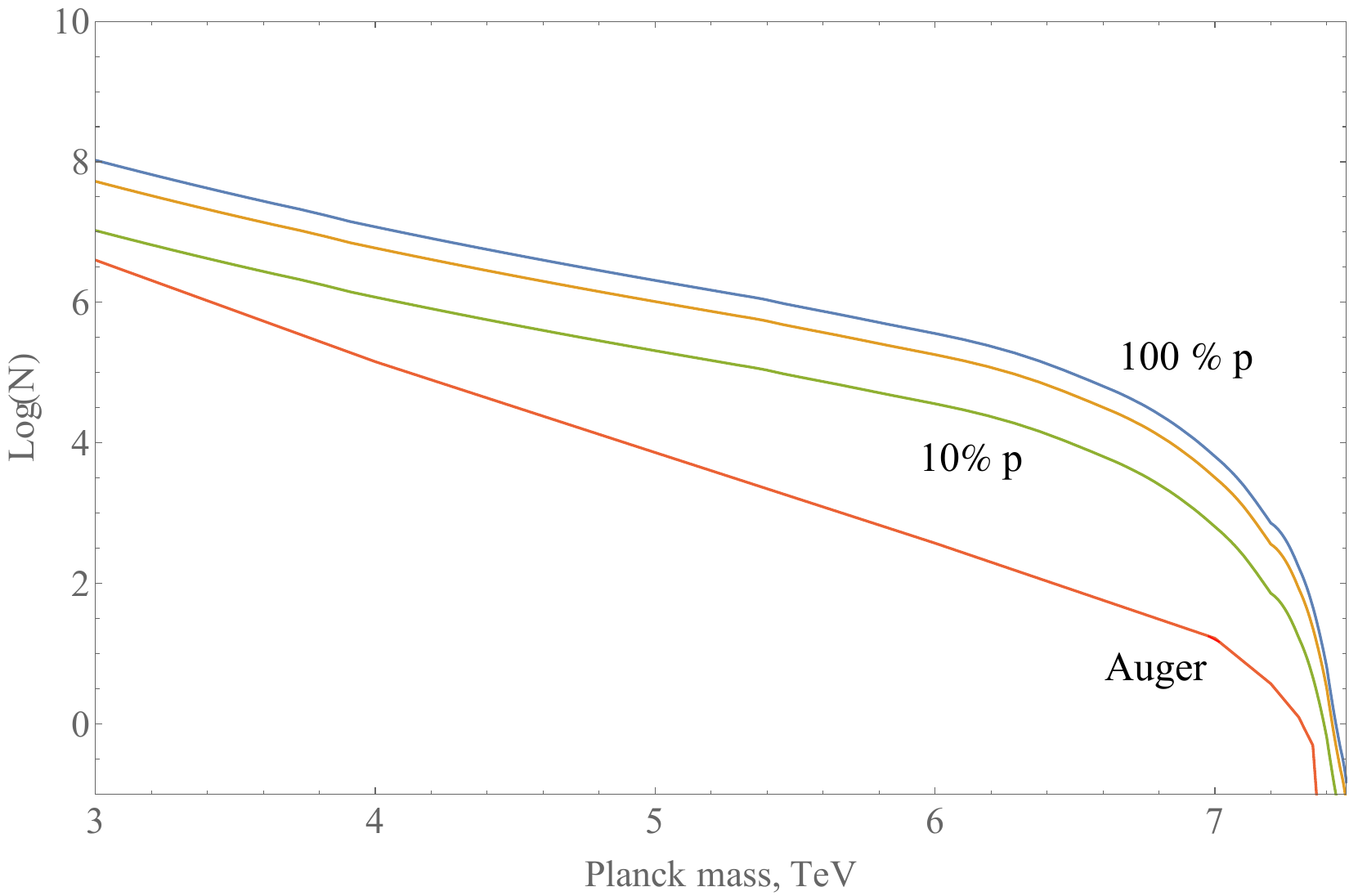}}
\ffigbox{\caption{The amount of the neutral black holes, stopped by a neutron star with a radius 10~km~ during $10^{10}$ yr, the cosmic ray energy is $E<5\cdot 10^{19}~$eV, for the different cosmic ray composition: $100\%$ p, $50\%$ p, $10\%$ p}\label{6f}}
{\includegraphics[height=4cm, width=6cm]{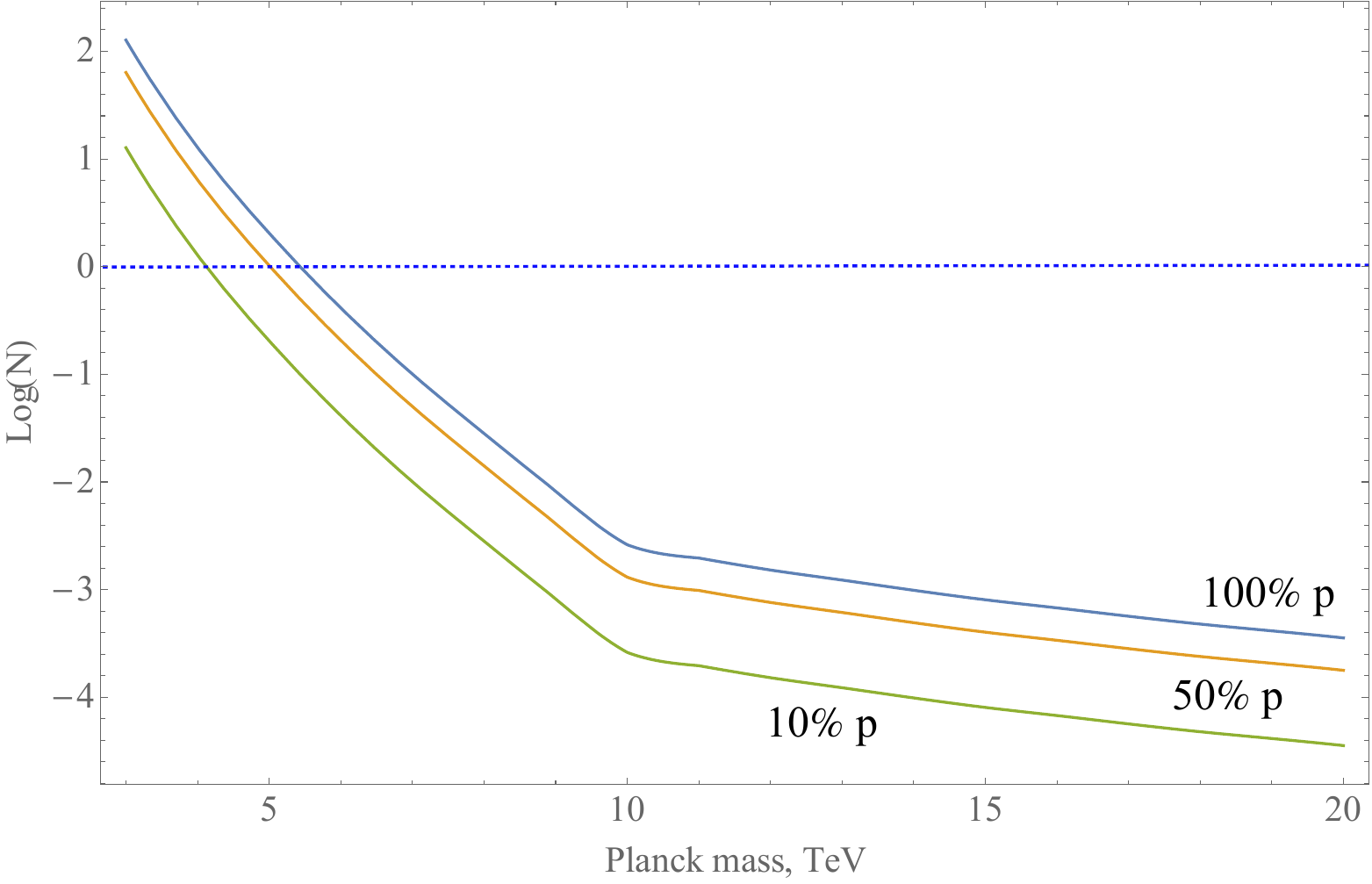}}
\end{floatrow}
\end{figure*}

The total number of the black holes, produced on a surface of a white dwarf of radius $R$ during the time $t$, is $N_{BH} = 4\pi R^2 \Omega t\cdot \phi_{BH}$, where the solid angle $\Omega = 2\pi$.
In order to calculate the whole number of the black holes captured during the lifetime of the white dwarf, we have to limit the region of the integration in the Eq. (\ref{5}) by the condition (\ref{3}) and account for a decrease of the column density $\delta$ for non-zero values of the angle of incidence of the cosmic ray $\alpha$. The latter is made using the dependence $\delta (\alpha)$ (Fig. 1 from \cite{Giddings:2008gr}). The overwhelming majority of the trapped black holes is given by the small $\alpha$, so we conservatively take $\delta = 0.8 \delta_{WD}$ (decreasing the value of $0.9\delta_{WD}$ by $10\%$ due to the possible systematic error),~ $1-\cos\alpha < 0.01$. The calculated number of the black holes stopped ($t=10^9~$yr, $R=5600~$km) is plotted as a function of $M_6$ in Fig. \ref{5f}, for the Auger data $A(E)$ and for the $100\%,~ 50\%,~ 10\%$ proton fraction in the cosmic rays. The calculation that uses Auger data on the composition sets the lower bound on the number of the black holes, because cosmic rays with the small atomic number $A$ (e.g. protons) give the main contribution to the black hole production, while increasing $A$ (till $A=56$ of iron) yields the number of black holes by orders of magnitude smaller. Thus, considering the mean value of $A$, we get the conservative estimates of the production. One can see that in the theories with $M_6<7.3~$TeV there is at least one black hole that is stopped by the white dwarf. One can see also that, due to the stopping condition, this constraint on $M_6$ would not improve significantly, even if we knew the precise 
cosmic ray composition. The composition that yields the biggest number of the black holes trapped (the most optimistic scenario) is $100\%$ p fraction and it gives only $M_6<7.4~$TeV.

\subsection{Constraints from neutron stars}
Now let us consider a neutron star. As the high energy cosmic rays cannot reach its surface due to the magnetic screening, we examine another, less effective mechanisms of the black hole production: 
\begin{itemize}
\item the production on baryons during the lifetime of the Universe. We conservatively assume that the cosmic ray flux is constant till the redshift $z=1$ (in reality it is supposed to increase with redshift). We also do not take into account the processes in the early Universe, which are quite model-dependent, considering redshifts $z<1$. The flux of the black holes can be obtained from Eq. (\ref{5}), substituting parameter $b$ for the following value:
\begin{equation}\label{cosm}
\begin{split}
\hat{b}=\int\limits_1^0 n(z) dt(z) = \hspace{45mm}& \\
  \int\limits_0^1 n_0 (1+z)^3 \frac{dz}{H_0 (1+z) \sqrt{\Omega_M(1+z)^3 + \Omega_\Lambda}}
= \quad & \\ 
\int\limits_0^1 \frac{n_0 (1+z)^2 dz}{H_0 \sqrt{\Omega_M(1+z)^3 + \Omega_\Lambda}},&
\end{split}
\end{equation}
where $n_0 = 2\cdot 10^{-7}~$cm$^{-3}$ is the current baryon density, $H_0 = 68 \cdot \frac{\text{km}/\text{s}}{\text{Mpc}}$ is the Hubble constant, $\Omega_M = 0.31$, $\Omega_\Lambda = 0.69$ (see \cite{Planck:2015klp}). In order to obtain reliable constraints, we set $E_{max} = 5\cdot 10^{19}~$eV in Eq. (\ref{5}), considering the cosmic rays with the energies below the Greisen-Zatsepin-Kuzmin(GZK) limit\cite{Greisen1966,Zatsepin1966}. For these energies the energy loss length for protons is more than 1~Gpc. We have also checked that explicit account for non-uniform distribution of extragalactic baryons \cite{Bi1997} very weakly affects our estimate. Eq. (\ref{cosm}) gives the value $\hat{b}=4.6\cdot 10^{21}~\text{cm}^{-2}$.
\item the production in binary systems of a neutron star and a red giant: cosmic rays hit the giant and produce black holes which then impinge on the neutron star (in the case of neutron stars we consider the neutral black holes, so they are not affected by the magnetic field); $\hat{b}=1/\sigma_{NN} = 10^{25}~\text{cm}^{-2}$, however $t\equiv~$'full coverage equivalent' $\lesssim 30~$Myr (see \cite{Giddings:2008gr}, Appendix H),
\item the production on interstellar medium ($\hat{b}=n_H$, where $n_H \sim 10^{21}~$cm$^{-2}$ -- average column density of hydrogen in the galaxy).
\item the production on the Central Molecular Zone of our Galaxy \cite{Morris1996}, for which $\hat{b}=nL=6\cdot 10^{22}~\text{cm}^{-2}$. However no long-lived neutron stars have been observed yet in this zone: maximal ages of the observed ones are $t=10^4 - 10^5~$yr.
\end{itemize}

The first three mechanisms give comparable fluxes of the black holes. The largest flux is achieved in the second mechanism, 6 times higher than in the first one and 30 times higher than in the third. The fourth mechanism produces flux that is 4-5 times lower. It is worth noting that the last mechanism will be the most constraining, if old millisecond pulsars are detected in the Central Molecular Zone. We use the first mechanism for our estimates, because the second one is more model-dependent and yields large systematic errors. The number of the black holes stopped by the neutron star ($t = 10^{10}~$yr, $R = 10~$km) is plotted as a function of $M_6$ in Fig. \ref{6f} in case of the different fractions of protons in the cosmic rays. The maximum $M_6$, that leads to more than one black hole stopped, is 4.1~TeV for $10\%$ proton composition, $5.0~$TeV for $50\%$ proton composition and $5.4~$TeV in case of $100\%$ proton composition. In this calculation we limit the energy of the cosmic rays from above by $5\cdot 10^{19}~$eV, so the most optimistic case of $100\%~$p composition is quite possible. Worse result, with the maximum $M_6$ around 3.5~TeV, is given by the mechanism of the cosmic rays (we take $A/Z \sim 2$) hitting the surface of the neutron star with the minimum magnetic field $B = 7\cdot 10^7~$G. The energy of the cosmic rays in this case is smaller than $5.9\cdot 10^6~$TeV due to the condition (\ref{4}); this energy is not sufficient for the production of the black holes with large masses. The magnetic field can be neglected at the poles, however at the high energies the surface area that can be reached there  is too small and the flux is less than in the case of the mechanisms studied above. In general, one can see that the constraints on $M_6$ are worse in the case of neutron stars than in the case of white dwarfs. However,  detection of the old neutron stars in the Central Molecular Zone and data on the composition of the cosmic rays can improve the existing constraints from neutron stars and make them the most robust.

\subsection{Case of the charged black holes}
Till now we considered the neutral black holes. The charged black holes will be stopped in a white dwarf for the whole range of energies due to electromagnetic interactions. In order to get the number of the black holes stopped during the lifetime of a white dwarf, we have to do the same calculations, as in the case of the neutral black holes, however ignoring the inequality (\ref{3}). For $100\%$ proton composition of the cosmic rays the number of the stopped black holes exceeds $6.6\cdot 10^4$ till the Planck mass of $14~$TeV. For this and bigger values of the Planck mass the production of black holes at the 100 TeV collider is zero, see Fig. \ref{2f}. Thus, the theories without the mechanism of Schwinger discharge yield more than one black hole trapped in a white dwarf during its lifetime, if the fraction of protons in the cosmic rays at energies from the process threshold $5\cdot 10^{18}~$eV to $5\cdot 10^{19}~$eV exceeds  $1.5\cdot 10^{-5}$, which is actually the case, according to the results of  Auger  \cite{Aab:2015bza} and  Telescope Array  \cite{Abbasi:2014sfa}.

According to the accretion theory, a micro black hole will accrete a neutron star very fast relative to the lifetimes of the known neutron stars (in our case accretion time $t_{acc} \lesssim 5.3\cdot (M_6/M_0)^2~$ min $\thinspace <1.5~$days). A white dwarf in our case will be destroyed in $t = 10^2\cdot (M_6/M_0)^2~$yr$~< 4\cdot 10^4~$yr, that is also negligible in comparison with the observed lifetimes of the white dwarfs. Thus, astrophysical observations constrain the theories in question and forbid the production of the charged stable micro black holes at the 100~TeV collider. The production of the neutral stable micro black holes is forbidden in the theories with Planck mass smaller than 7.3~TeV.
 
\subsection{Constraints from astrophysical neutrinos}
\begin{figure*}
\caption{Upper bound on the amount of the black holes, produced due to hypothetical high energy neutrino collisions with a neutron star with a radius 10 km during 10 million years and stopped by it}\label{8f}
\includegraphics[height=6cm]{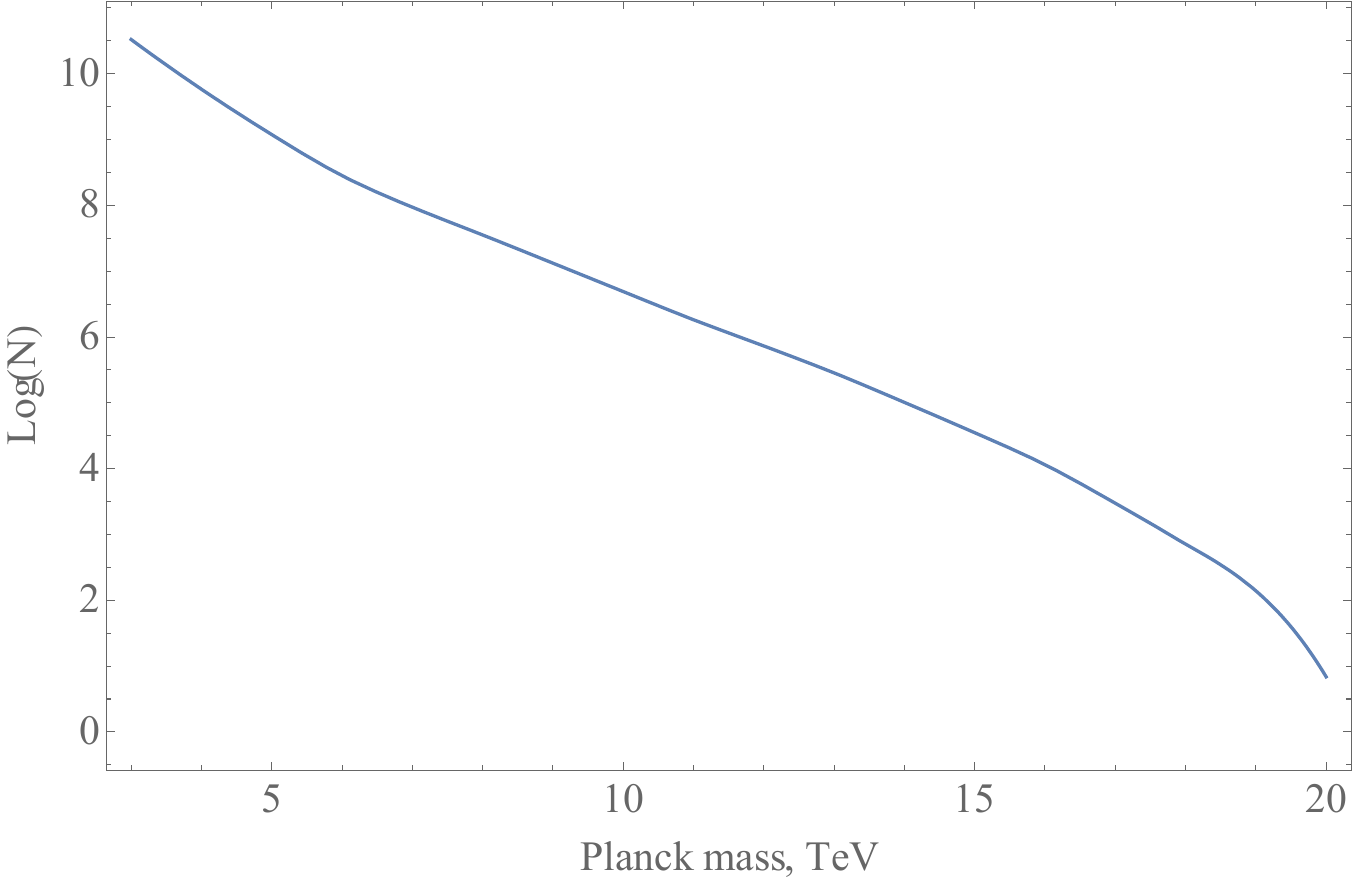}
\end{figure*}

One has to mention that robust astrophysical constraints hypothetically can be obtained from the mechanism of high energy neutrinos hitting the surface of a neutron star. The known effect of the decay of high energy neutrino in the large magnetic field into an electron and a W-boson \cite{Borisov:1985ha,Erdas:2002wk,Kuznetsov:2010sn,Satunin:2014toa} does not influence these constraints. Indeed, in the article \cite{Erdas:2002wk} it was found that above the process threshold $E_{th} \sim 2.2 \cdot 10^{16} \cdot (B_{cr} / B)~$eV an asymptotic neutrino absorption length is $l \sim 1.1 \cdot (B_{cr} / B)^2 \cdot (10^{16}~\text{eV} / E)~$m, where $E$ is the neutrino energy, $B$ is the magnetic field and $B_{cr} = 4.4 \cdot 10^{13}~$G. Thus the effect of neutrino decay becomes significant only for the neutron stars with the magnetic fields about $10^{12}~$G and higher. Observation of the neutron stars with the small magnetic fields, in particular observation of a neutron star with the magnetic field $B \sim 7\cdot 10^7~$G \cite{Manchester:2004bp}, suggests that this effect can be neglected in our study. We constrain the number of the black holes trapped inside a neutron star with the small magnetic field, using the upper limit on the flux of single-flavour high energy neutrinos \cite{Aab:2015bza}:
\begin{equation}
N (E) < 6.4\cdot 10^{-9} \cdot E^{-2} \text{ GeV cm}^{-2} \text{ s}^{-1} \text{ sr}^{-1},
\end{equation}
valid for the energies $E = 0.1 - 25$~EeV. The resulting bounds on the number of the black holes trapped per ten million years are presented in Fig. \ref{8f}. The calculation was made using a formula analogous to the Eq. (\ref{5}):
\begin{equation}
\begin{split}
N_{BH} = 2\pi & S t \int\limits_{E_{min}}^{E_{max}} 3 N(E) dE \times \\
&\sum\limits_{i} \int\limits_{x_{min}}^1 dx \cdot f_i(x) \cdot \frac{\sigma '  (\sqrt{s'})}{\sigma '  (\sqrt{s'})+\sigma_{tot} (E)},
\end{split}
\end{equation}
where $s' = 2m_p E x$, $E_{min} = M_{min}^2 /(2m_p y^2)$ -- the threshold of black hole production, $x_{min} = E_{min}/E$, $E_{max} = 25~$EeV, $\sigma_{tot}$ is the full neutrino-nucleon inelastic cross section, see \cite{Gandhi:1998ri}. Inelasticity was conservatively taken to be $y = 0.5$. The number of neutrino flavours was accounted for by the multiplier 3 in the integrand.

Thus, the mechanism of high energy neutrinos hitting the surface of a neutron star can provide robust constraints as soon as a high energy ($E>10^5$~TeV) neutrino flux is detected experimentally. Now the highest energy of the detected astrophysical neutrino does not exceed $10^4~$TeV \cite{Aartsen:2013bka}.

\section{Conclusion}

\begin{figure*}
\caption{The constraint from white dwarfs on the number of the neutral black holes, trapped inside the Earth}\label{9f}
\includegraphics[height=6cm]{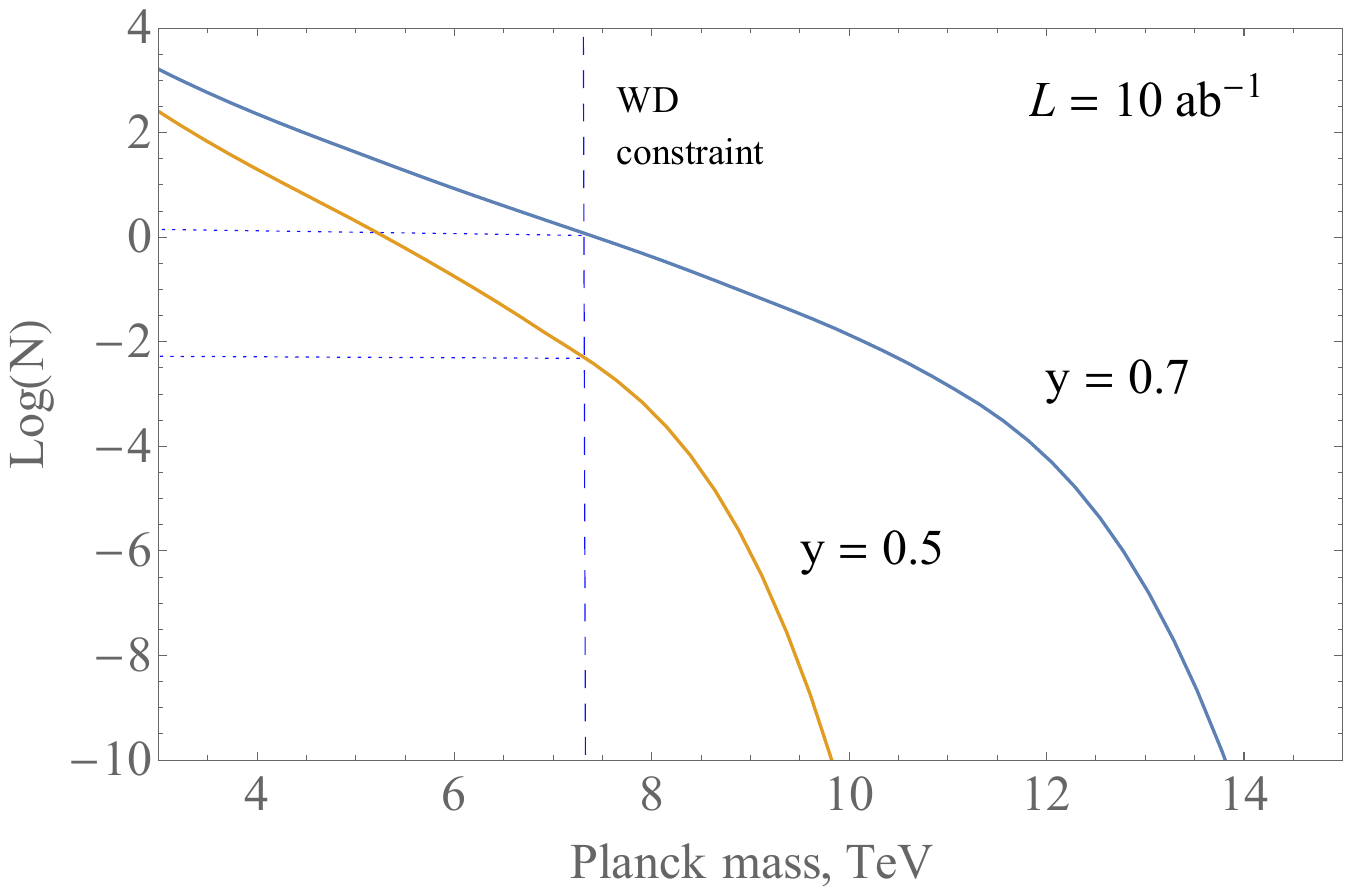}
\end{figure*}

In this article we have studied the phenomenology of the models with extra dimensions in absence of the Hawking radiation in order to conduct an independent observations-based check of safety of the proposed 100~TeV collider.
The models with more than 6 dimensions always yield Earth's accretion times larger than the lifetime of the Solar system. A theory with five dimensions could be consistent with the existing experimental constraints on the size of extra dimensions and yield accretion times smaller than the lifetime of the Solar system only with a fine-tuning of the warp-factor, given by the inequalities (\ref{6}). The calculation of the number of the micro black holes that would have been produced in the future 100~TeV collider with the integrated luminosity $L = 10~$ab$^{-1}$ and the astrophysical constraints from the observational data on the lifetime of white dwarfs and cosmic ray composition suggest that it is possible to exclude the production of the charged stable micro black holes already. As it is shown in Fig. \ref{9f}, the case of the neutral black holes, while broadly addressed for most $D$ and mass values, leaves some loophole, which can be closed with further cosmic ray data (e.g. on the neutrino spectrum and cosmic ray composition) or astrophysical observations.

\begin{acknowledgements}
  The authors would like to thank  P.~Satunin,~S.~Troitsky, I.~Tkachev and the anonymous referee for numerous valuable discussions and comments on the manuscript. The question of the safety of a future collider, in its relation to cosmic-ray studies, was asked by J.-R. Cudell at the Solvay meeting ``Facing the scalar sector''. The work of the authors was supported by  the Russian Science Foundation grant 14-12-01340. This research has made use of NASA's Astrophysics Data System.
\end{acknowledgements}

\end{document}